\documentstyle[sprocl]{article}

\input{epsfig1.sty}

\def\drop#1{}

\def\cut#1{}
\newcommand{\equ}[1]{~Eq.~(\ref{#1})}

\newcommand{\sla}{\raise.15ex\hbox{$/$}\kern -.8em}

\newcommand{\half}{\frac 1 2}

\newcommand{\bra}{\langle}\newcommand{\ket}{\rangle}
\newcommand{\be}[1]{\begin{equation}\label{#1}}
\newcommand{\ee}{\end{equation}}
\newcommand{\ba}[1]{\begin{eqnarray}\label{#1}}
\newcommand{\ea}{\end{eqnarray}}

\title{A Quantum Approach to Stock Price Fluctuations}
\author{Martin Schaden$^\dagger$}
\address{New York University, 4 Washington Place, New York, New York 10003}
\date{\today}
\begin{document}
\maketitle
\begin{abstract}
\noindent A simple quantum model explains the L{\'e}vy-unstable
distributions for individual stock returns observed by ref.[1].
The probability density function of the returns is written as the
squared modulus of an amplitude. For short time intervals this
amplitude is proportional to a Cauchy-distribution and satisfies
the Schr{\"o}dinger equation with a non-hermitian Hamiltonian. The
observed power law tails of the return fluctuations imply that the
"decay rate", $\gamma(q)$ asymptotically is proportional to $|q|$,
for large $|q|$. The wave number $q$, the Fourier-conjugate
variable to the return $x$, is interpreted as a quantitative
measure of "market sentiment". On a time scale of less than a few
weeks, the distribution of returns in this quantum model is shape
stable and scales. The model quantitatively reproduces the
observed cumulative distribution for the short-term normalized
returns over 7~orders of magnitude without adjustable parameters.
The return fluctuations over large time periods ultimately become
Gaussian if $\gamma(q\sim 0)\propto q^2$. The ansatz
$\gamma(q)=b_T\sqrt{m^2+q^2}$ is found to describe the positive
part of the observed historic probability of normalized returns
for time periods between $T=5$~min and $T\sim 4$~years over more
than 4~orders of magnitude in terms of one adjustable parameter
$s_T=m b_T\propto T$. The Sharpe ratio of a stock in this model
has a finite limit as the investment horizon $T\rightarrow 0$.
Implications for short-term investments are discussed.
\end{abstract}
\footnotetext{$^\dagger$ Email address: m.schaden@att.net}

\section{Introduction}
The quest for a quantitative statistical description of stock
prices began more than a century ago with Bachelier's
thesis\cite{Ba00} in which he described price movements by a
random walk. Although Bachelier's random walk has since been
modified, the effort to model stock price movements by a
stochastic process continues. Computerized historic financial
records have recently made a high-precision statistical analysis
of short-term returns possible. For time scales that are less than
a few weeks it is increasingly difficult to reconcile this wealth
of new information with the conjecture that price fluctuations are
described by a stochastic process with independent and identically
distributed increments ({\it iid}).

To estimate the probability density function (pdf),
$p_T^{(i)}(x_T^{(i)})$, for the return $x_T^{(i)}$ (the change in
the logarithm of the price $S^{(i)}$ of the stock $(i)$ over a
time period $T$),
\be{defx}
x_T^{(i)}:=\ln[S^{(i)}_{t+T}/S_t^{(i)}]=\ln[S^{(i)}_{t+T}]-\ln[S^{(i)}_t]\
,
\ee
the frequency of such changes was analyzed in ref.[1], using
historical records that typically cover several years. Reducing
the time horizon $T$ to minutes offered the advantage of
suppressing the effects of a changing macro-economic environment
while at the same time improving the statistics.
{\vskip0.1truecm\epsfig{figure=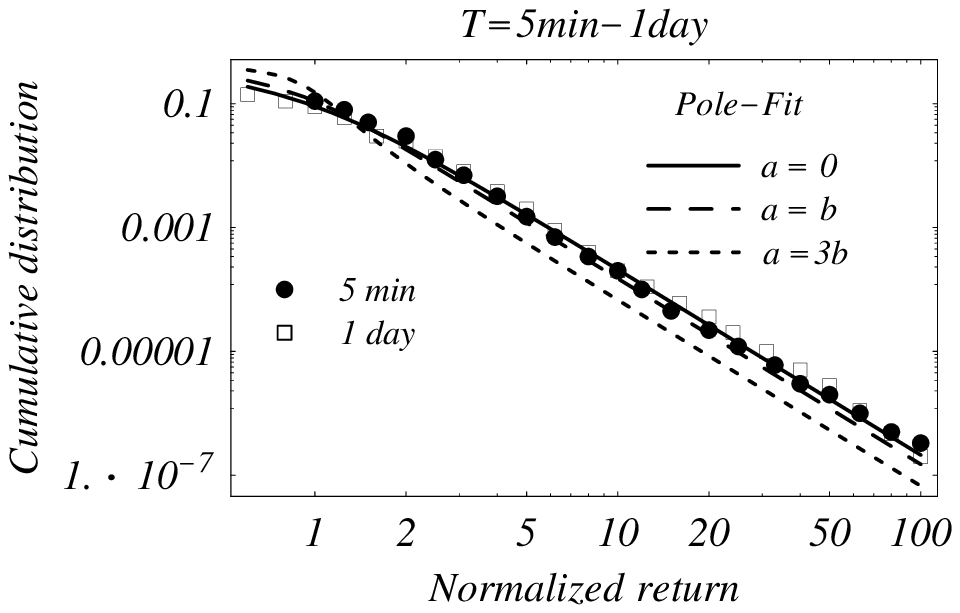,height=7.0truecm}\nobreak

{\small\noindent Fig.~1. Fits to the observed average cumulative
distribution for short-term normalized returns. The historic
probabilities for time periods of $T=5{\rm ~minutes}$ (dots) and
$T=1{\rm ~day}\sim 390{\rm ~minutes}$ (squares) are from the
analysis of ref.[1]. Only the positive tails of the slightly skew
distributions are shown. The lines correspond to cumulative
distributions of the form given in\equ{normcum} for three
different ratios, $r=(a/b)=0,1,3$, of the parameters $a$ and
$b$.}} \vskip 5pt

The analysis of ref.[1] revealed that the distribution of returns
for different stocks and for different time intervals $T<2$~weeks
are all shape-similar and exhibit a somewhat unexpected power law
fall off. More precisely: if $\mu_T^{(i)}$ denotes the mean and
$v_T^{(i)}$ the standard deviation of the returns $x_T^{(i)}$ over
a period $T$ for the stock $(i)$, the cumulative distribution of
the normalized returns $g$,
\be{cumdis}
N_T^{(i)}(g)={\bf E}[x_T^{(i)}-\mu_T^{(i)}>v_T^{(i)} g]\ ,
\ee
was found\cite{St99} to not depend on the stock $(i)$ nor on the
time horizon $T$ for $5{\rm ~min}<T<2{\rm ~weeks}$. The mean,
$\mu_T^{(i)}$, and standard deviation, $v_T^{(i)}$, thus were
found to be the only statistically significant characteristics of
the historical distribution of returns of any individual stock on
time scales up to a few weeks. {\it Scaling} in the following
denotes the fact that the distributions of the {\it normalized}
returns are very similar. Although this would also be true for a
random walk or any other L{\'e}vy stable process, it is somewhat
unexpected that non-Gaussian return distributions with a {\it
finite} variance are shape-similar over vastly different time
horizons $T$, for very different companies and in very different
economic environments.

The empirically observed scaling of the distributions permits a
significant improvement of the statistics by averaging over the
distributions of the normalized returns of many individual
stocks\cite{St99},
\be{avcum}
N_T(g):=\frac{1}{{\#\rm ~stocks}} \sum_{i=1}^{{\#\rm ~stocks}}
N_T^{(i)}(g)\ .
\ee
One thus can reliably estimate the probability for return
fluctuations that are about $100$~times larger than average.

The shortest time horizon investigated by ref.[1] is $T=5{\rm
~min}$. The analysis in this case follows the stocks of the
1000~US companies with the largest market capitalization over a
2-year period from January 1994 to December 1995. The average
cumulative probabilities for the normalized $5$-minute returns
found by ref.[1] is reproduced in the log-log plot of Fig.~1. Also
shown is the average cumulative distribution of the normalized
daily returns. The latter were extracted\cite{St99} from the
records of stock prices for about 16,000~individual companies
(binned by market capitalization) over the entire 35-year interval
1962-96.

The interested reader is referred to the original analysis in
ref.[1] for further details. But I would like to emphasize that
the observed scaling of the distributions of individual companies
(and of groups of companies with different market capitalization)
is crucial to the interpretation of the analysis. The average
normalized cumulative distribution of\equ{avcum} otherwise would
differ qualitatively from the normalized distribution of any
individual stock. Without scaling, the error in the mean
normalized distribution of\equ{avcum} would be large and the data
base of about $4\times 10^4$ events per company would not allow
any conclusions about the frequency of events that occur only a
few times out of $10^{7}$.

Fig.~1 shows that the averaged distributions for different time
intervals $T<1{\rm ~day}$ also scale extremely well. It is
remarkable that normalized data sets for time intervals that
differ by a factor of about $78\sim (1{\rm ~trading~day})/5{\rm
~min}$ and involve different companies in different economic
periods should show no statistically significant difference.

Apart from this shape-similarity over vastly different time scales
and many different companies, the most striking feature of the
empirical probability distribution is the pronounced power law
over $\sim5$~orders of magnitude,
\be{scaling}
N_T(g)\propto g^{-\alpha}\ ,\ \ {\rm for}\ 2<g<100\ {\rm and}\
T<1{\rm ~week}\ .
\ee
The exponent $\alpha$ empirically is close to $3$. For $T=5$~min,
the best estimate for $\alpha$ is $3.10\pm 0.03$ for the positive
tail and $2.84\pm 0.12$ for the negative tail of the observed
distribution\cite{St99}. For short time intervals $T<2~{\rm
weeks}$ the distributions are only slightly skewed, but this
feature becomes more pronounced in the data with increasing $T$.
The observed asymmetry of the distributions could be caused by a
number of factors, such as the discreteness of prices or
bankruptcy regulations, that all tend to mainly distort its
negative tail. We do not model such effects and for the purpose of
this investigation replace the empirical distribution by a
symmetric one with the same positive part.

The empirical estimates of the power law exponent are well outside
the region for L{\'e}vy-stable distributions, which would
require\cite{Ma00} that $0<\alpha\leq 2$. The stability of the
power law tails with an exponent $\alpha\sim 3$ refute Bachelier's
conjecture\cite{Ba00} and its generalizations. The observed
distributions cannot result from an {\it iid} process. The pdf for
$T=390{\rm ~min}\sim1{\rm ~day}$ otherwise would have to differ
considerably in shape from the pdf for $T=5{\rm ~min}$; the
$390/5=78$-fold convolution of the latter is rather close to a
Gaussian pdf and not at all shape-similar to itself.

The observed auto-correlation time is of the order of a few
minutes\cite{Ma00}only. Memory effects that could explain the
persistence of power law tails over days therefore probably are of
higher order. Although GARCH-processes\cite{Bo86} in principle
model such correlations, it is quite difficult to obtain a
shape-stable temporal evolution with the observed power law in
this manner\cite{Le01}. A GARCH-process furthermore depends on
parameters that determine the shape of the marginal distribution.
In order to reproduce the observed {\it scaling}, one would have
to contend that these shape degrees of freedom somehow are
strongly correlated and very similar for different companies,
macro-economic conditions, etc.

Power law tails with exponents $\alpha\sim3$ have also been
observed for the cumulative distributions of market
indices\cite{Go99} and of commodity prices\cite{Ma02}. Although
the existence of fat tails was recognized early on, an economic
explanation of the power law with an exponent $\alpha\sim 3$ has
only recently been proposed\cite{Ga02}. In this micro-economic
analysis, the asymptotic power law reflects the distribution and
trading behavior of the largest investors in a stock, such as
mutual- and/or pension- funds. However, such an {\it asymptotic}
analysis cannot, by itself, explain the actual magnitude of the
tails nor the observed scaling of the distributions.

If one ignores the problem of temporal stability, the distribution
of returns on a stock (including the power law tails) for any
fixed (short) time interval $T$ can be modelled by a subordinated
stochastic process. Clark\cite{Cl73} originally proposed a
subordinated stochastic process in his attempt to explain daily
returns on cotton futures. He contended that such a process would
give a better description of cotton futures than Mandelbrot and
Taylor's suggestion\cite{Ma67} of a L{\'e}vy stable process. Only
a limited amount of data ($\sim 2000$ prices) on the daily closing
of cotton futures was analyzed and the issue of shape-stability of
the distribution of returns over longer (or shorter) time
intervals than 1~day was not addressed. The daily returns on
cotton futures indeed are better reproduced by a subordinated
stochastic process than by a L{\'e}vy stable one, mainly because
the latter can give a leptokurtic distribution with finite
variance.

More recently, the Variance Gamma process\cite{Ma98}, a
subordinated stochastic process with finite kurtosis, has been
proposed to value options. A subordinated stochastic process with
finite variance but infinite kurtosis that reproduces the observed
distribution of returns on a stock for any {\it fixed} time
interval $T<2$~weeks is constructed in the appendix. However, this
{\it iid} process with finite variance converges rapidly to its
Gaussian fixed point and does not explain the shape-stability of
the observed distribution on different time scales $T$.

In short, the observed price fluctuations pose at least three
theoretical challenges:
\begin{itemize}
  \item[1.] To explain the scaling of the distribution of short-term
  returns of many individual stocks over $\sim 7$~orders of magnitude.
  \item[2.] To explain the power law behavior with $\alpha\sim 3$ of the
  tails of the cumulative distribution for the returns.
  \item[3.] To explain the apparent temporal shape-stability of the
  distribution for the returns over time horizons from a
  few minutes to a few weeks and the slow convergence
  to a Gaussian distribution on much longer time scales.
\end{itemize}

The functional form of the observed normalized distribution of
short-term returns is accurately reproduced by the ansatz of the
next section. In section~3 the temporal evolution of the {\it
amplitude} (the square root of the pdf) is found to be naturally
shape-stable in a quantum model and the linear evolution operator
(effective Hamiltonian) that reproduces the observed temporal
evolution for $T<1{\rm ~day}$ is obtained. Section~4 discusses and
interprets a slightly modified effective Hamiltonian. It
semi-quantitatively describes the temporal evolution of the
distribution of returns over time scales up to four years.
Section~5 explores some of the implications of this quantum
mechanical description and the results are summarized in
section~6.

\section{Power Law Tails with $\alpha=3$ in Quantum Finance}
The recently proposed quantum description of financial
markets\cite{Sc02} offers a surprisingly simple and transparent
explanation for the observed distributions. The exponent
$\alpha\sim3$ for the power law tails, in particular, is quite
natural in this framework and is temporally {\it stable}.

$\alpha=3$ in the cumulative distribution of\equ{cumdis} implies
the asymptotic behavior,
\be{asypdf}
p_T(x^2\sim\infty)=- \left.\frac{\partial}{\partial x}
N_T(x/v_T)\right|_{x^2\sim\infty}\propto x^{-4}\ ,
\ee
of the pdf of a representative stock's return. The essence of a
quantum description is that the pdf is interpreted as the squared
magnitude of a (possibly complex) amplitude $\phi_T(x)$,
\be{ampl}
p_T(x)=|\phi_T(x)|^2\ .
\ee

If the temporal evolution (and thus the pdf) is invariant under
the trans\-for\-ma\-tion$^\#$
\footnotetext{$^\#$This is an (approximate) "parity"-symmetry for
reflection about the mean $\mu$.} $x-\mu \rightarrow \mu-x$, the
amplitude either is symmetric or antisymmetric in $x-\mu$. Since
the probability of the mean return, $p_T(\mu)$, does not vanish,
$\phi_T(x)$ in this case is symmetric and a function of
$(x-\mu)^2$ only. The asymptotic behavior of the pdf
in\equ{asypdf} implies
\be{asyphy}
\phi_T(x^2\sim\infty)\propto x^{-2}\ ,
\ee
for the asymptotic behavior of the amplitude.  Since $\phi_T(x)$
falls off like a power law  for large values of $x$ with an
exponent that is (close to) a negative integer, we model the
amplitude by a rational function, i.e. a function that is analytic
in the whole complex plane apart from a finite number of poles.
The position and strength of the poles in this case constitute the
set of parameters that describe the pdf of an individual stock.
The observed approximate scaling implies that the distribution of
an individual stock's returns to first approximation is specified
by its variance and mean. Our ansatz therefore should not involve
too many poles. Every additional parameter is related to
additional$^*$\footnotetext{$^*$With more than two poles, the pdf
can, for instance, be skewed.} statistical properties that in
general will break the scaling and distinguish between the {\it
shapes} of the distribution of returns of individual stocks. An
amplitude with a single pole corresponds to a pdf with infinite
variance. The observed finite variance thus requires an amplitude
with at least two poles and that the sum of residues must vanish.
The amplitude then asymptotically falls off at least as fast as
in\equ{asyphy}. Note that only asymptotic power laws with an
exponent $\alpha$ that is an {\it odd integer} can be modelled in
this fashion and that an odd exponent $\alpha>3$ would require
more than two poles$^\dagger$\footnotetext{$^\dagger$For
$\alpha$'s that are not odd integers, the amplitude would have
branch cuts}.

The simplest rational amplitude corresponding to a pdf of finite
variance therefore has two poles of opposite strength. It
automatically has the asymptotic behavior of\equ{asyphy}. The
requirement that the amplitude is a symmetric function of $x-\mu$
further constrains the position of the two poles to $z =\mu\pm(a+i
b)$. One thus is led to consider the ansatz,
\be{mero}
\phi_T(x)=\frac{{\cal N}}{(x-\mu)^2-(a+i b)^2},
\ee
for the amplitude. The real parameters $\mu$, $a$ and $b>0$
generally will depend on the stock, economic era and the time
interval $T$.
 Normalizing the pdf determines the constant ${\cal N}$ up to an
 irrelevant phase,
\be{norm}
|{\cal N}|^2=(a^2+b^2)\frac{2 b}{\pi}\ .
\ee
Note that the pdf is not normalizable at $b=0$. The normalized pdf
corresponding to the ansatz of\equ{mero} is,
\be{pdfansatz}
p_T(x)=\frac{2 b (a^2+b^2)}{\pi(((x-\mu)^2+b^2-a^2)^2+4 a^2 b^2)}.
\ee
The three real parameters, $\mu$, $a$ and $b>0$ on which it
depends are related to the mean, variance and the curvature at the
mode of the density. The variance of $p_T(x)$ is just the square
of the distance of the complex poles from the mean,
\be{var}
v^2=|z-\mu|^2=a^2+b^2 \ .
\ee
The second (independent) parameter is the ratio $r=a/b$, or,
equivalently, the phase of $z-\mu$. The curvature of the pdf at
$x=\mu$ is
\be{curvature}
\rho=\left.\frac{\partial^2}{2\partial x^2}
p_T(x)\right|_{x=\mu}=\frac{4b}{\pi v^4}\frac{a^2-b^2}{a^2+b^2} .
\ee
$\rho$ is negative for $a^2<b^2$ and the pdf in this case peaks at
$x=\mu$. For $a^2>b^2$ the pdf is double-humped with two distinct
maxima at $x_{\rm max}=\mu \pm \sqrt{a^2-b^2}$. $\rho v^3$ is a
function of the ratio $r=a/b$ only.

The distribution for the normalized returns that corresponds to
the pdf of\equ{pdfansatz} is,
\be{normcum}
N_T(g)=\frac{1}{2\pi}\left[ \pi+\arctan\left(\frac{a-v
g}{b}\right)-\arctan\left(\frac{a+v g}{b}\right)+\frac{b}{2
a}\ln\left(\frac{1+[\frac{a-v g}{b}]^2}{1+[\frac{a+ v
g}{b}]^2}\right)\right]\ .
\ee
With\equ{var}, $N_T(g)$ is seen to depend on the shape parameter
$r^2=(a/b)^2$ only. For large and small values of $g$, $N_T(g)$
has the expansions,
\ba{asN}
N_T(g\sim\infty)&\sim& \frac{2b}{3\pi v}
g^{-3}\left[1+\frac{3\pi\rho v^4}{10 b }
g^{-2}+O(g^{-4})\right]\nonumber\\
N_T(g\sim 0)&\sim& \frac{1}{2}-\frac{2 b}{\pi v}g-\frac{\rho
v^3}{3} g^3+O(g^5)\ .
\ea
[The relative correction to an asymptotic power law $\propto
g^{-3}$ is never much more than $6/(5 g^2)$ (or less than $1\%$
for $g>11$)]. The leading correction vanishes altogether for
$\rho=0$, or equivalently, $r=1$.

Fig.~1 shows distributions $N_T(g)$ of\equ{normcum} for ratios
$r=0,1,3$, together with the empirical data. Apart from the
overall change in the normalization of the tails, the parameter
$r$ qualitatively changes the shape of the distribution of
normalized returns near $g=0$. Note that the theoretical
distributions, by construction, all have the same power law tails
with an exponent $\alpha=3$, but that the normalization of the
tails of the distribution for $r=3$ is off by a factor of $\sim
20$. Any asymptotic analysis that predicts only the
exponent\cite{Ga02}, by itself, therefore cannot provide an
explanation of the observed distribution of returns.

There is no convincing reason why parameters like $r$ should be
the same for different companies and/or economic eras or different
time intervals $T$. Large variations in $r$ would break the
observed shape similarity of the distributions. Within the
accuracy of the analysis of ref.[1], the ratio $r=a_T/b_T$ does
not appear to depend strongly on $T$ and should be similar for
most stocks. The empirical data prefers small $r<1$ and is
consistent with $r=0$, i.e. with an amplitude whose two poles have
similar real parts. In this case the amplitude describing a mean
return of $\mu_T$ with standard deviation $v_T=b_T$ has two
complex conjugate poles of opposite strength located at
$z_T=\mu_T+i v_T$ and $z_T^*=\mu_T-i v_T$. This is the minimal
number of independent parameters the distribution of returns of an
individual stock can depend upon. The normalized probability
distribution corresponding to this special case is the solid line
in Fig~1, a student t-distribution with 3~degrees of freedom,
\be{specialpdf}
p_T(x)=\frac{v_T}{2\pi}\left|\frac{1}{x-z_T}-\frac{1}{x-z_T^*}\right|^2=
\frac{2 v_T^3}{\pi((x-\mu_T)^2+v_T^2)^2}\ .
\ee
It fits the data exceedingly well. For $T<2$~weeks, the returns of
a stock in this sense are statistically described by the path the
pole $z_T$ takes in the complex half-plane with ${\rm Im}\,
z_T>0$. The path begins at the origin $z_{T=0}=0$ and long term
investments are encouraged if the standard deviation $v_T$ is a
concave function of the mean return $\mu_T$.

\section{Quantum Stability}
The good description of the empirical data by an ansatz for the
pdf that is the square of the modulus of a rational amplitude
could be considered fortuitous and in itself does not require a
quantum model. But a quantum model of the dynamics does explain
the apparent stability of the shape of this distribution over
vastly different time intervals without the need to explicitly
model memory effects. It also casts some doubt on the notion that
stock market fluctuations are predictable if they do not follow an
{\it iid} process.

Quantum dynamics primarily describes the temporal evolution of the
amplitude -- the evolution of the corresponding pdf follows from
the relation in\equ{ampl} and is non-linear. Stochastic processes
that give L{\'e}vy-unstable distributions of the type
\equ{pdfansatz} (with the correct power law tails) over an
extended period of time\cite{Le01,Bo01,Ma97} tend to be arbitrary
in the sense that other distributions may almost equally well have
been obtained. They either depend on quite a few carefully
adjusted parameters\cite{Le01,Bo01} or are directed by a process
whose fractal dimension would have to be a time independent
characteristic of financial markets\cite{Ma97}. Unlike for
physical phenomena, such "hidden" characteristics of financial
markets are expected to slowly change over time and the
distributions thus would depend on the economic era. The ansatz
of\equ{mero} for the amplitude on the other hand turns out to be
absolutely shape stable under a relatively simple and quite
natural quantum dynamics.

To better see this, consider the Fourier-transform of the
amplitude of\equ{mero} with the normalization\equ{norm}.  Contour
integration about the simple poles of the ansatz at $z=\mu_T
\pm(a_T+i b_T)$ with $b_T>0$ is elementary and gives the
Fourier-transform $\widetilde\phi_T(q)$,
\be{fourieramp}
\widetilde\phi_T(q):=\int_{-\infty}^\infty \phi_T(x) e^{i q x}dx
=\sqrt{2\pi b_T}\, e^{i(q\mu_T +|q|a_T+i |q|b_T)}\ .
\ee
For very short times, when the mean and variance of the
distribution are small, $\tilde\phi_T(q)$ essentially is constant
over a wide range of wave numbers $q$. In this case the
corresponding amplitude\equ{mero} for the return is well localized
about $x=\mu$, i.e. the expected log-price. The variance of the
pdf grows with the time horizon if $a_T$ and/or $b_T$ grow in
magnitude. As noted before, the cumulative distribution of the
normalized returns remains strictly the same only if the
parameters $a_T$ and $b_T$ grow proportionally and the ratio
$r=(a_T/b_T)$ does not depend on the time interval $T$.

The temporal evolution of a state $|\varphi\ket$ in the Hilbert
space of a quantum model is generated by some effective
Hamiltonian operator $\hat H_{\rm eff}(t)$,
\be{evolution}
|\varphi\ket_T={\bf T}e^{-i\int_0^T dt\hat H_{\rm
eff}(t)}|\varphi\ket_0\ ,
\ee
where the symbol ${\bf T}$ denotes time ordering of the
exponential factors (redundant if, as in the present case,  the
Hamiltonian operators for different times commute). To conserve
probability in a complete Hilbert space, $\hat H_{\rm eff}(t)$
necessarily would have to be hermitian and states would evolve by
a unitary rotation. However, if $\hat H_{\rm eff}(t)$ describes
only the evolution in a subspace of the Hilbert space -- such as
that spanned by a single share -- probability can be "lost" to the
complementary part of the full Hilbert space.  $\hat H_{\rm
eff}(t)$ in this case is not hermitian.

Comparing\equ{evolution} with\equ{fourieramp} suggests that the
effective Hamiltonian for the time evolution of a share is
diagonal in the Fourier-conjugate basis of $|q\ket$-states, with
matrix elements,
\be{Hmat}
\bra q|\hat H_{\rm eff}(t)|q'\ket=2\pi\delta(q-q')(-q\dot \mu_t
-|q|\dot a_t-i|q|\dot b_t)\ ,
\ee
where the dot is shorthand for the derivative with respect to
time, i.e.  $\dot\mu_t=d\mu_t/dt $, etc.

In the language of particle physics, $\widetilde\phi_T(q)\propto
\bra q|\varphi\ket_T$ is proportional to the wave-function of a
(massless) particle whose frequency $\omega(q)$ and decay rate
$\gamma(q)$ are both proportional to the magnitude of its
wave-number $q$. If $r=a/b$ does not depend on time, one can
interpret $b_T$ as proportional to the "proper" time of a stock,
i.e. as the monotonically increasing parameter that characterizes
the evolution of the stock's price distribution. $b_T$ need not be
proportional to physical time or even to trading time. The proper
time of a share could be proportional to some monotonically
increasing quantity that is relevant to the actual trading
dynamics of the stock, such as the overall number of trades or the
number of traded shares. We will further examine this issue in
section~5.

From the available data one in fact cannot conclude that the
frequency $\omega(q)$ and decay rate $\gamma(q)$ of a share are
proportional. As Fig.~1 shows, the data implies only that the term
proportional to $a_T$ is small compared to $b_T |q|$. Since $a_T$
and $b_T$ are the strengths of two commuting terms of the
evolution operator, the contribution $\omega(q) a_T$ may simply be
negligible. $a_T=0$ is a trivial fixed point of the evolution that
reproduces the empirical data rather well. [Note that this would
not be the case if one could not separate the effects of the two
parameters in $\hat H_{\rm eff}$.] Assuming $|a_T|\ll b_T$ in this
sense is a "natural" approximation to the evolution that does not
require any fine tuning.

The time dependent proportionality constant $\sqrt{2\pi b_T}$ in
the amplitude $\widetilde\phi_T(q)$ of\equ{fourieramp} ensures
that the pdf remains normalized at all times when $b_T>0$. It does
not appear in\equ{evolution} and the norm of the state
$|\varphi\ket_T$ decays with $T$ if $b_T$ increases with time. The
time dependent normalization factor arises because the effective
"one-particle" Hamiltonian $\hat H_{\rm eff}$ describes the
temporal evolution of a particular share of a stock. We are
therefore computing the probability of a return $x$, if a certain
share traded at time $t$ is again traded at time $t+T>t$. The
probability that any particular share is again traded (for any
price) after a time interval $T$ is just the normalization
${_T}\bra\varphi|\varphi\ket_T=(2\pi
b_T)^{-1}{_0}\bra\varphi|\varphi\ket_0$ of the state$^{\$}$
\footnotetext{$^{\$}$We will see in section~5 that it is conceptually
more appropriate to include a factor $2\pi b$ in the relation of
\equ{ampl} between the conditional probability density and the
square of the amplitude, instead of normalizing the square of the
amplitude.} $|\varphi\ket_T$ in\equ{evolution}. The probability
that a particular share is traded after a time interval $T$ thus
decreases like $1/b_T$, but the probability for a particular
share's return under the condition that it is again traded after a
time $T$ of course remains normalized for all $T$. [The
calculation is analogous to that of the probability of a decaying
particle's change in position {\it if} it is observed after a time
$T$.]

\section{Ultimate Convergence to a Gaussian Fixed Point}
The exponent $\alpha\sim3$ and the apparent temporal shape
stability of the distribution thus are linked by the fact that
quantum theory primarily describes the temporal evolution of the
amplitude, rather than of the transition probability. Although the
evolution of an amplitude of Cauchy type such as the one
in\equ{mero} is shape stable, there are other possible fixed
points. Of some interest is the Gaussian one, because one does
expect the pdf to eventually approach a Gaussian distribution.
Note that the Fourier-transform of a Gaussian amplitude is a
Gaussian and that a Gaussian amplitude implies a Gaussian pdf.

The pdf $p_T(x)$ of individual stocks empirically resembles a
Gaussian distribution only after several years. The results of the
analysis of ref.[1] for $T=16{\rm ~days}$ to $T=1024{\rm ~days}$,
or up to approximately $4{\rm ~trading~years}$, are shown in
Fig.~2. The slow rate of convergence to a Gaussian fixed point
suggests that it may be due to macro-economic and other factors
such as stock splits, buy-backs, etc., that are irrelevant for the
short-term dynamics of a stock. Note that the positive and
negative tails of the cumulative distribution do not approach the
Gaussian fixed point at the same rate\cite{St99}. Although the
approximation by a symmetric distribution becomes questionable on
long time scales, it nevertheless should be possible to
consistently describe the crossover from one (almost) fixed point
to another within a single effective quantum model. For simplicity
and because it is a fixed point and consistent with the short-term
data, we consider only the case $a_T=0$, i.e. a stock whose
evolution is completely dominated by its "decay rate" $\gamma(q)$.
{\vskip0.1truecm\epsfig{figure=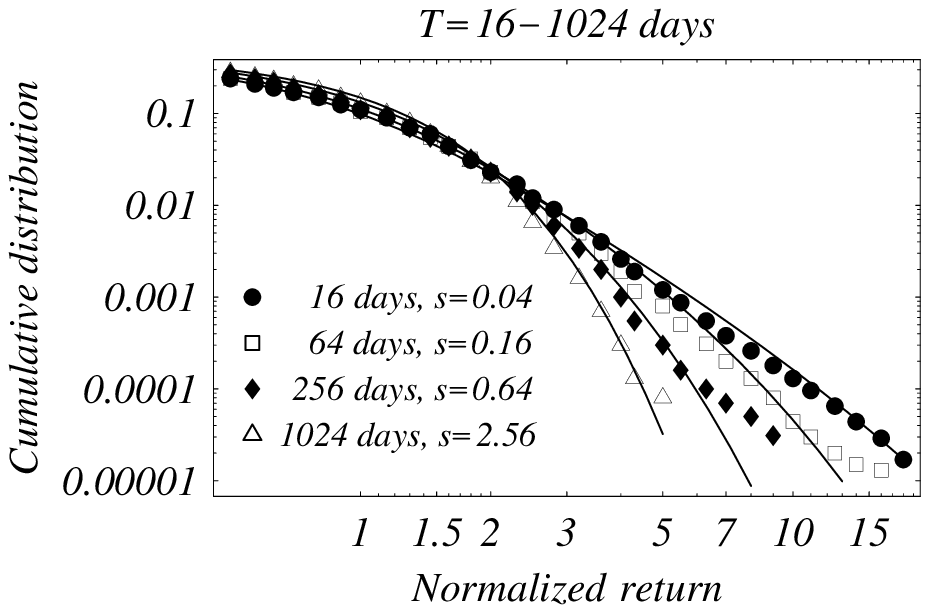,height=7.0truecm}\nobreak

{\small\noindent Fig.~2. The cumulative distribution for the
normalized returns over time intervals from $T=16$ to $T=1024$
trading days. The empirical probabilities are from the analysis of
ref.[1]. The full lines are distributions that correspond to the
extended ansatz of\equ{extend} that best fit the data. They depend
on the value of the parameter $s=b m$ only.}} \vskip 5pt

For long time horizons, the variance of the pdf and thus of the
amplitude are large. The Fourier-conjugate amplitude
$\widetilde\phi_T(q)$ therefore is concentrated about $q\sim 0$,
and the approach to a Gaussian fixed point can be ensured if the
function $\gamma_0(q)=|q|$ in\equ{fourieramp} is modified to one
where the cusp at $q=0$ is replaced by a smooth quadratic
dependence on $q$. It was argued previously\cite{Sc02}, that the
effective decay rate $\gamma(q)$ should be a quadratic function of
$q$ in the long wavelength limit, $q\sim 0$. The proposed
modification for small values of $q$ will not affect the
short-term power law behavior over a wide range of $x$ if
$\gamma(q)$ for $|q|\sim\infty$ approaches the function $|q|$
sufficiently rapidly. Replacing $|q|$ by the upper branch of a
hyperbola with these asymptotes thus is a possibility,
\be{omega}
|q|\rightarrow \gamma_m(q)\propto\sqrt{q^2+m^2}-m\ .
\ee
The parameter $m$ controls the transition between the Gaussian and
power law regimes. For $m=0$ one recovers the previous case,
whereas the density essentially is Gaussian for all but the
shortest times when $m$ is large. For $m>0$ the ultimate fixed
point thus is Gaussian, but the convergence to it is slow for
small $m$. The particular choice for the interpolating function
in\equ{omega} is ad hoc, but a square-root dependence of the decay
rate on the wave-number can be the result of diagonalizing a
(anti-)hermitian $2\times 2$ matrix -- such square-roots are quite
generic for quantum systems near level crossings.
{\vskip0.3truecm\epsfig{figure=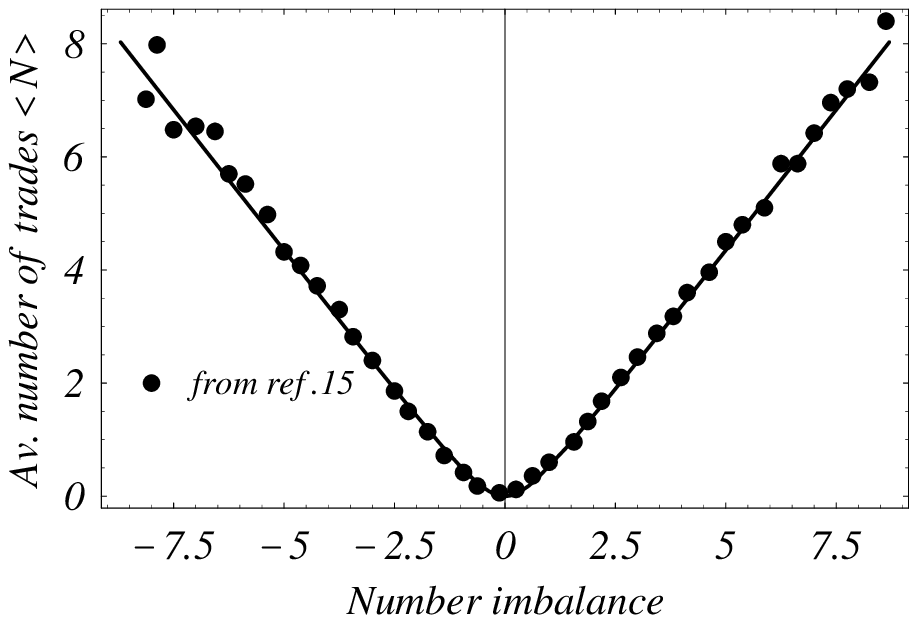,height=7.0truecm}\nobreak

\small\noindent Fig.~3. The conditional expectation $<N>_\Phi$ of
the number of trades for a given number imbalance $\Phi$. The data
points are reproduced from Fig.~3a) of the analysis of ref.[15].
The solid line represents the function $\sqrt{\Phi^2+0.49}-0.7$.}
\vskip 5pt

Empirical support for the particular form of $\gamma_m(q)$
in\equ{omega} is provided by the analysis in ref.[15] of how the
mean number of trades $\bra N\ket_{\Phi_T}$ in a time interval of
$T=15$~min depends on the imbalance between the buyer- and
seller-initiated trades. Whether a trade is buyer- or seller-
initiated was determined\cite{Pl01} from the prevailing quote
shortly before execution of the trade using the procedure of
C.~Lee and M.~Ready\cite{Le91}; the imbalance $\Phi_T$ then is the
difference in the number of buyer- and seller- initiated trades in
the time interval. The data points in Fig.~3 are the results of
ref.[15] from the analysis of the trading records of the 116 most
frequently traded US stocks during the 2-year period 1994-1995.

If one assumes that the average volume of shares per trade does
not depend strongly on $\Phi_T$, the observations of ref.[15] can
be interpreted and understood within the quantum model. Because
the time interval of $T=15$~min is relatively short,
$<N>_{\Phi_T}$ in this case is proportional to the "decay" rate
$\gamma_m(q)$ of a share in state $|q\ket$. The observations in
Fig.~3 indeed appear to lie on a symmetric hyperbolic curve.
Furthermore, there appears to be little trading activity at
vanishing imbalance $\Phi_T\sim 0$. It therefore is tempting to
assume that the wave number $q$ and the imbalance $\Phi_T$ are
roughly proportional. This interpretation would associate the
market dynamics captured by the imbalance $\Phi_T$ with the
Fourier conjugate variable to the return $x_T$ (up to a
proportionality constant). It has potentially interesting and
observable consequences in a quantum model. The pdf of $\Phi_T$ in
this case is related to the pdf of $x_T$: the corresponding
amplitudes are Fourier conjugate to each other (up to a scale).
The amplitude $\widetilde\phi_T(q)$ of\equ{fourieramp} implies
that for time intervals $T<2$~weeks the normalized distribution of
$\Phi_T$ is ${\bf E}[\Phi>g v_{\Phi_T}>0]\sim \half
e^{-g/\sqrt{2}}$, where $v_{\Phi_T}\propto 1/b_T$ is the standard
deviation of $\Phi_T$. The model thus predicts that the
probability of a large imbalance in the trades falls off
exponentially, rather than like a power law. [Note that
$v_{\Phi_T}\propto 1/b_T$ is one version of Heisenberg's
uncertainty relation. It here implies that periods with large
fluctuations in the imbalance should show relatively small
fluctuations in the returns, and vice versa.]

Returning to the temporal evolution of the distribution of the
returns, the modification of\equ{omega} leads to an amplitude of
the form$^{\&}$\footnotetext{$^{\&}$We assume that $a_T\ll b_T$
and neglect any dependence on the parameter $a_T$. This
approximation is consistent with the short-term results.},
\be{extend}
\phi_T(x;m)=\frac{1}{{\cal N}}\int_0^\infty \frac{dq}{\pi} \cos(q
x) \exp[-b\sqrt{q^2+m^2}]=\frac{m b}{\pi {\cal N}}
\frac{K_1(m\sqrt{x^2+b^2})}{\sqrt{x^2+b^2}} ,
\ee
where $K_1(z)$ is the modified Bessel function of the third kind
with the asymptotic behavior
\be{asympK}
K_1(z\sim 0)\sim 1/z,\ \ \
K_1(z\sim\infty)\sim\sqrt{\frac{\pi}{2z}}e^{-z}\ .
\ee
Normalization of the transition probability requires that,
\be{normm}
|{\cal N}|^2=\int_0^\infty \frac{dq}{\pi} e^{-2
b\sqrt{q^2+m^2}}=\frac{m}{\pi} K_1(2 m b)\ .
\ee
For a mean return $\mu$, the pdf for the return on a stock in this
model then becomes,
\be{pdfm}
p_T(x;\mu,b,m)=|\phi_T(x-\mu;m)|^2=\frac{m b^2}{\pi K_1(2 m
b)}\frac{K_1^2(m\sqrt{(x-\mu)^2+b^2})}{(x-\mu)^2+b^2}\ ,
\ee
and depends on three, possibly $T$~dependent, parameters
$\mu_T,b_T$ and $m_T$. The variance of the returns is
\be{varm}
v_T^2=\frac{b^2}{m K_1(2 m b)}\int_0^\infty \frac{q^2 dq}{q^2+m^2}
e^{-2 b\sqrt{q^2+m^2}}\ .
\ee
For long time horizons ($m b\gg 1$), the limiting density is
Gaussian with a variance $v^2(m b\gg 1)\sim b/(2 m)$. For short
time intervals ($m b\ll 1$) the variance to first approximation
does not depend on $m$, $v^2(m b\ll 1)=b^2\left(1-\pi m b
+\dots\right.$.

Since the variance of normalized returns is unity by definition,
the cumulative distributions for the normalized returns depends on
the parameter combination $s=m b$ only. As shown in Fig.~2, this
extended ansatz qualitatively and to some extent even
quantitatively reproduces the observed\cite{St99} positive tails
of the (average) cumulative distributions of the normalized
returns. The distributions for time periods of $T=16,64,256$ and
$T=1024$~trading days correspond to values of the parameter $s=m b
=0.04,0.16,0.64$ and $s=2.56$. We obtain that
\be{timescale}
s_T=m_T b_T\approx T/(400 {\rm days})\ .
\ee
This determination of the time scale unfortunately is not very
accurate and depends among other things on our assumption that all
stocks are characterized by the same $m_T$ (see below). The
approach to a Gaussian fixed point of the negative tails is even
slower\cite{St99}. The fact that we are approximating skewed
distributions by symmetric ones may partly explain some of the
discrepancies visible in Fig.~2.

Systematic deviations also arise if the individual stock
distributions scale less than perfectly for large $T$. If the
parameter $m_T$ depends on the stock, the scaling among companies
is broken on large time scales and the "lightest" stocks (the ones
with the smallest $m_T$) in this case dominate the average
of\equ{avcum} at large normalized returns. The observed flattening
out of the tails of the averaged cumulative distributions for
$T>16$~days in this case could be due to some "lighter than
average" stocks.

The overall quality  of this one-parameter fit down to observed
cumulative probabilities of $10^{-5}$ is encouraging. The fact
that $s_T$ is approximately proportional to the trading time $T$
furthermore suggests that the effective Hamiltonian could be time
independent. For sufficiently long time intervals one would hope
this to be the case for any parameterization of the evolution --
most monotonically increasing quantities that might be relevant
for the evolution of a stock's price eventually do become
proportional to the physical- (and to the trading-) time.

\section{Quantum Interference and Short Term Risk and Return}
For long time horizons $T$ one expects that the mean return
$\mu_T$ and the variance $v_T^2\sim b_T/(2 m_T)$ of the returns
are both proportional to $T$. We found that the time dependence of
the distribution fits the empirical one if $s=m_T b_T$ is
approximately proportional to $T$. Taken together, this would
imply that on time scales  of $T\sim$~years,
\be{largeT}
\mu_T\propto b_T\propto T, \ \ {\rm and}\ \ m_T=m\sim {\rm
const.}\ \ .
\ee

If one assumes time homogeneity of the returns, the quantum model
in fact completely specifies the $T$-dependence of the parameters
$\mu_T$, $b_T$ and $m_T$ for all $T$. Assuming that the
distribution of historic returns of a stock over a time interval
$T$ does not depend on the initial time $t$ is not very reasonable
for single companies, but perhaps is tenable for short time
intervals $T$ once companies are grouped by their market
capitalization -- the procedure followed in ref.[1]. The
conditional probability density $p_T(y|x)$ that the stock has a
return $y-x$ in the time interval $T$ in this case is,
\be{transp}
\left.p_T(y|x)\right|_t=p(y,t+T|x,t)\propto |G(y,t+T;x,t)|^2\ ,
\ee
where $G(y,t+T;x,t)$ is the transition {\it amplitude} that a
particular share traded at time $t$ for a log-price $x$ is traded
at time $t'=t+T>t$ for a log-price between $y$ and $y+dy$. [The
proportionality constant in general depends on $T$ and $t$ and is
obtained from $1=\int p(y,t+T|x,t)\,dy$.]

The transition amplitude in the present model
is$^\ddagger$\footnotetext{$^\ddagger$The states $|x\ket$ and
$|y\ket$ are normalized eigenvectors of the hermitian operator
$\hat x$ that corresponds to the log-price. The scalar
product\cite{Baym} with the previously introduced state $|q\ket$
is $\bra q|x\ket=e^{i q x}$.},
\ba{Green}
G(y,t+T;x,t)&&=\bra y| {\bf T}e^{-i\int_t^{t+T} \hat H_{\rm
eff}(\xi) d\xi}|x\ket\nonumber\\&&\hskip-3em=\int_{-\infty}^\infty
\frac{dq}{2\pi} \exp\left[i(x-y)q +i\mu_T(t) q
-b_T(t)\sqrt{q^2+m_T^2(t)}\right]\ .
\ea
An effective Hamiltonian that is diagonal in the wave number
implies that $G(y,t+T;x,t)$ and $p_T(y|x)$ depend on the return
$y-x$ only. If the returns were homogeneous in time,
$G(y,t+T;x,t)$ would not depend on the initial time $t$. This is
an oversimplification of the dynamics and we will consider the
possibility that the parameters in\equ{Green} vary slowly with the
initial time, i.e. that the returns are homogeneous on time scales
of the horizon $T$.

The definition of\equ{Green} and the completeness of the basis of
$|y\ket$-states gives Trotter's formula for the transition {\it
amplitudes},
\ba{convol1}
G(z,T_1+T_2+t;x,t)&=&\bra z| {\bf T}e^{-i\int_{t+T_1}^{t+T_1+T_2}
\hat H_{\rm eff}(\xi) d\xi}e^{-i\int_t^{t+T_1} \hat H_{\rm
eff}(\xi) d\xi}|x\ket\nonumber\\
&&\hskip-5em =\int_{-\infty}^{\infty} dy\bra z| {\bf
T}e^{-i\int_{t+T_1}^{t+T_1+T_2} \hat H_{\rm eff}(\xi)
d\xi}|y\ket\bra y|{\bf T}e^{-i\int_t^{t+T_1} \hat H_{\rm eff}(\xi)
d\xi}|x\ket\nonumber\\
&&\hskip-5em=\int_{-\infty}^{\infty} dy\ G(z,T_1+T_2+t;y,T_1+t)
G(y,T_1+t;x,t)\ .
\ea
\equ{convol1} is the mathematical expression of Huygens' principle
describing the propagation of waves. Using\equ{Green} and
performing the integral over the intermediate log-price $y$, one
finds that for\equ{convol1} holds only if,
\ba{timedep}
\mu_{T_1+T_2}(t)&=&\mu_{T_1}(t)+\mu_{T_2}(t+T_1)\nonumber\\
b_{T_1+T_2}(t)&=&b_{T_1}(t)+b_{T_2}(t+T_1)\nonumber\\
m_{T_1+T_2}(t)&=&m_{T_1}(t)=m_{T_2}(t+T_1)=m(t)\ .
\ea
If the parameters depend little on the initial time (at least on
time scales $T_1$ of interest), $\mu_{T_2}(t+T_1)\sim
\mu_{T_2}(t)$ and $b_{T_2}(t+T_1)\sim b_{T_2}(t)$. The solution
to\equ{timedep} in this case is that $\mu_T(t)$ and $b_T(t)$ are
both proportional to the time interval $T$,
\be{soltimedep}
\mu_T(t)\sim T\dot\mu(t) ,\ \ b_T(t)\sim T\dot b(t),\ \ \
m_T(t)=m(t)\ .
\ee
For notational clarity, the possible dependence of parameters on
the initial time $t$ will again be suppressed in the following.

On a time scale of years, the $T$-dependence of the parameters is
consistent with\equ{largeT} and implies that both, the mean and
the variance $v_T^2\sim T\dot b/(2m)$ of long-term returns is
proportional to $T$. On long time scales, Sharpe's ratio ($\dot r$
is the risk-free return rate),
\be{Sharpesratiolong}
\lambda_T=\frac{\mu_T-\dot r T}{v_T}\propto T^{1/2} \ \ {\rm for}
\ \ T m\dot b \gg 1
\ee
is a monotonically increasing function of the investment horizon
that is similar to that predicted by an {\it iid} process. On
short time scales, the Sharpe ratio of the present model does not
increase as fast as for an {\it iid} process. Since the standard
deviation  $v_T\sim T\dot b$ is itself proportional to $T$,
Sharpe's ratio in particular does not vanish for $T\rightarrow 0$,
\be{Sharpesratioshort}
\lambda_0=\lim_{T\rightarrow 0}\frac{T(\dot\mu-\dot
r)}{v_T}=\frac{\dot\mu-\dot r}{\dot b} > 0\ .
\ee
If returns were described by an {\it iid} process, short-term
trading strategies would be extremely risky and should not find
rational investors. The persistence of day-trading and of
"excessive" trading by institutional investors\cite{Od99} suggests
that short-term strategies may not be prohibitively risky. The
prolonged existence of such phenomena is more plausible if the
specific risk, $1/\lambda_T$, remains finite for $T\rightarrow 0$.
Although the risk per unit of excess return does rise, short-term
investments are not penalized excessively by the price dynamics of
the quantum model. [Short term investments nevertheless can be
very expensive due to transaction costs.]
{\vskip0.1truecm\epsfig{figure=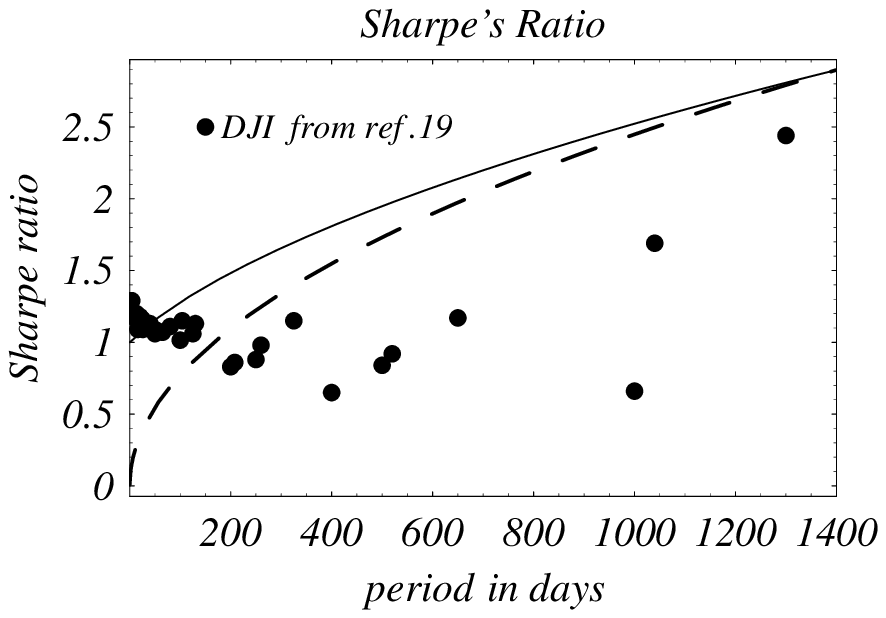,height=7.0truecm}\nobreak

{\small\noindent Fig.~4: The time dependence of Sharpe's ratio.
The points are from ref.[19] and give the observed Sharpe ratio of
the Dow-Jones Industrials index. The solid line is the Sharpe
ratio the quantum model predicts for an individual stock. The time
scale has been calibrated by\equ{timescale} and corresponds to the
one of Fig.~2. The overall normalization of the Sharpe ratio of
the stock was adjusted to the same order of magnitude as the
index. For comparison, the dashed line gives the time dependence
$\lambda(T)\propto \sqrt{T}$ of the Sharpe ratio that an {\it iid}
process would predict. }} \vskip 5pt

The Sharpe ratio of the historic returns on the Dow-Jones
Industrials index (DJI) in fact is fairly constant for periods up
to $T<100$~days\cite{Pe94} (and perhaps even declines somewhat).
The results from Peters' analysis of the DJI index are shown in
Fig.~4 and compared with the time dependence of the Sharpe ratio
of a single representative stock in the quantum model. Although
Peters studied the Sharpe ratio of an index rather than of
individual stocks, his findings do indicate that the specific risk
does not become prohibitively large as the investment period is
shortened.

To better see what a constant Sharpe ratio may mean to investors,
compare the following two short-term investment strategies
($0\leq\alpha\leq 1$):
\begin{itemize}
\item[S1] An amount $\alpha W$ is invested in asset $A$ and
the remainder, $(1-\alpha) W$, is invested in asset $B$ over a
period $T$.
\item[S2]  The full amount $W$  is invested in
asset $A$ for a time $\alpha T$ and for the remaining time,
$(1-\alpha)T$, the full amount is reinvested in asset $B$.
\end{itemize}
If the mean and the standard deviation of the returns on the
individual assets $A$ and $B$ are uncorrelated and proportional to
the investment period, both strategies carry the same short-term
risk (as measured by the standard deviation of the overall
returns) and have the same expected overall return. The
equivalence of the two strategies does not depend on the ratio
$\alpha$ nor on the standard deviations and mean returns of the
individual assets$^{**}$\footnotetext{$^{**}$The overall return
and risk of either strategy does of course depend on the returns
on the individual assets and in general is optimal for a
particular value of $\alpha$ only.}. If the Sharpe ratio
approaches a finite value for $T\rightarrow 0$, a short-term
investor therefore could be (almost) indifferent to choosing asset
diversification (strategy S1) or time diversification (strategy
S2). Since time diversification offers the option of choosing the
reinvestment time at a later point in time, it will often be
preferred for short-term investments. [Note that the transaction
costs incurred by both strategies could be similar and that time
diversification may only be penalized by the fact that
$\dot\lambda_0>0$.]

Less direct evidence for $b_T\propto T$ is provided by the
probability density at zero return, $p_T(0)=p_T(x|x)$.
\equ{specialpdf} implies that for $T<2$~weeks, $p_T(0)\sim
\frac{2}{\pi v_T}$ in this model. With $v_T\sim b_T\propto T$, the
probability density at zero return on short time scales therefore
is inversely proportional to the period, $p_T(0)\propto T^{-1}$.
$v_T\propto \sqrt{T}$ on the other hand would imply $p_T(0)\propto
T^{-1/2}$. Neither of these power laws corresponds very well to
$p_T^{\rm S\&P500}(0)\propto T^{-0.71\pm0.03}$ observed for the
probability density at zero return of the S\&P500
index\cite{Ma95}. But the stocks in an index are not perfectly
correlated. Diversification decreases the variance of the index
relative to the variance of the single-name stocks. Since
correlations tend to decrease over time, the probability density
at zero return on the index therefore increases relative to that
of a single-name stock. A slower decay over time of the
probability density at zero return for the index than for a single
stock thus is more plausible than a faster
one$^{\dagger\dagger}$\footnotetext{$^{\dagger\dagger}$The
probability density at zero return decays equally fast ($\propto
1/\sqrt{T}$) for the stock and the index in the (uncorrelated)
Gaussian case}. The prediction of the present model that
$p_T(0)\propto T^{-1}$ for an individual stock thus may well be
compatible with a probability density at zero return $p_T^{\rm
S\&P500}(0)\propto T^{-0.71\pm0.03}$ observed for the S\&P500.

One should emphasize that with the definition\equ{transp},
Trotter's relation\equ{convol1} in general is not compatible with
the relation for the transition probability densities of a Markov
chain,
\be{markovchain}
p_{T_1+T_2}(z|x)=\int p_{T_2}(z|y) p_{T_1}(y|x)\, dy\ ,
\ee
because the absolute square of a sum of terms in general is not
the sum of absolute squares of the individual terms. The
difference is known as quantum interference. A Gaussian transition
amplitude (and consequently a Gaussian transition probability
density) is an important exception where \equ{convol1} {\it and}
\equ{markovchain} both hold. The absence of interference effects
characterizes the (incoherent) classical limit of a quantum
system. Applying this criterion to the observed return
distributions, the classical description of equity markets only
becomes accurate on a timescale of years.

Note that\equ{markovchain} may be violated in quantum theory even
if the (quantum) process is memoryless and the Markov property
holds for the conditional probability densities, i.e. if
$p(z,t''|y,t';x,t;\dots)=p(z,t''|y,t')$ for all $t''>t'>t$. This
does not lead to any logical contradiction if price measurements
at intermediate times have a non-negligible effect on the
distribution. In a financial context this would means that trading
a share to determine its price at an intermediate time results in
a different distribution for the final outcome than if it were not
traded.

Although such back-coupling effects are well known and observable
on financial markets, it is generally difficult to quantitatively
include them in stochastic models. The quantum framework
incorporates "measurement"-effects in a consistent and tractable
(but perhaps rather narrow) fashion. There may be no need to model
the impact from trading separately.

\section{Summary}
The observed\cite{St99} cumulative distributions of the short-term
returns on single-name stocks are accurately modelled by pdf's of
the form in\equ{specialpdf}. These densities are the square of an
amplitude that is an analytic function of the returns apart from
two complex conjugate poles of opposite strength. The poles are
located at $z=\mu_T+i v_T$ and $z^*=\mu_T-i v_T$, where $\mu_T$ is
the mean return and $v_T$ is the standard deviation of the returns
over the time period $T<2$~weeks$^{\$\$}$\footnotetext{$^{\$\$}$It
is tempting to associate these singularities of the amplitude with
long and short positions in the stock}. This mathematically rather
concise description trivially accounts for the observed\cite{St99}
power law tails (with an exponent $\alpha=3$) and also implies
{\it scaling} between the distributions of different individual
companies.

The fact that the distribution of short-term returns with finite
variance approximately retains its shape over time scales
$T<2$~weeks rules out an {\it iid} process and is thus challenging
to describe stochastically. The temporal evolution of the
corresponding {\it amplitude}  on the other hand was found to be
given by the remarkably simple {\it linear} operator $\hat H_{\rm
eff}(t)$ of\equ{Hmat} that is proportional to the wave-number $q$.
[The temporal evolution of the corresponding pdf (the absolute
square of the amplitude) is not linear.] Since a dependence of the
effective Hamiltonian on fractional powers of the wave number is
unusual for quantum systems, $\alpha=3$ in this sense is a
"natural" exponent for distributions with stable power law tails.

The proposed quantum model gives a surprisingly quantitative and
transparent explanation of the observed shape and temporal
stability of the observed stock price fluctuations on short time
scales. As Fig.~1 shows, the model quantitatively reproduces the
average cumulative distribution observed for normalized 5~minute
and daily returns over $7$~orders of magnitude with no free
parameters.

By regulating the cusp of $|q|$ at $q=0$ and thus modifying the
non-hermitian part of the Hamiltonian to a quadratic function of
$q$ near $q=0$, the shape similarity of the evolution is broken
and eventual convergence to a Gaussian probability density over
time horizons of several months to years is assured. With the
modification of\equ{omega} the model qualitatively reproduces the
observed distributions for the normalized returns\cite{St99} down
to cumulative probabilities of $10^{-4}$ for time periods up to
$4$~trading years. As discussed in section~4, the deterioration in
the quality of the fit on long time scales may partly be due to
the skewness of the observed distributions, which is more severe
for longer time periods. It could also be due to some
"lighter-than-average" stocks if the scaling of the individual
stock distributions for $T>16$~days is less than perfect. Such
stocks would dominate the average of\equ{avcum} at high normalized
returns and explain the flattening out of the observed tails.

Comparing the prediction of this model for the "decay-", or
trading- rate of a share with the observed\cite{Pl01} dependence
of the trading rate on the imbalance between buyer- and
seller-initiated trades, suggest that the imbalance number of
ref.[15] is proportional to the wave-number. The imbalance in this
case is the Fourier conjugate variable to the return $x_T$ (up to
an overall scale). The quantum model then predicts a definite form
for the distribution of the imbalance number -- it in particular
should have exponential, rather than power law, tails.

By including sufficiently many "hidden variables", that is
factors, stochastic models may be able to reproduce the historic
observations with similar accuracy. Indeed, the (possible)
existence of hidden variables was at the heart of the early
critique of quantum theory.  For physical quantum phenomena, this
alternative explanation has only recently been rejected by
experimental verification of Bell's inequalities\cite{Bell}.

It is quite impossible to perform similar high-precision
experiments on equities in order to reject the existence of
"hidden variables". There in fact is no need to because it is
almost self-evident that the return on a stock depends on many
factors that have not been modelled. The question nevertheless is
not just one of having a more efficient description of the
dynamics. Unlike hidden variables describing physical phenomena,
the factors that influence the dynamics of a stock are expected to
change over time. It furthermore is not clear {\it how} economic
factors like the gross national product influence the value of any
given stock at any given point in time. The observed scaling of
the return distributions for various stocks in different economic
environments strongly suggests that all these "hidden" factors
find their expression in the mean return and the variance of the
returns. The dynamics that determines the shape of the return
distributions on the other hand must be self-consistent and
largely immune to the influence of "hidden" variables that are
specific to a company and the economic and political climate.

The conceptual advantage of a quantum description is that this
consistent framework describes the observed return fluctuations
rather well {\it without} separately modelling all the possibly
influential, time-dependent, and hard-to-measure factors. Contrary
to a purely descriptive "hidden variable" model, the consistency
of a quantum model leads to a number of predictions. Their
verification or falsification will ultimately decide whether this
is a useful approach.

{\bf Acknowledgement:} I would like to thank several members of
the applied mathematics department of the Courant Institute of New
York University for their hospitality and for organizing the very
informative and accessible seminar in mathematical finance where I
became aware of the empirical analysis discussed here. I am very
much indebted to Larry Spruch for his personal support.
\appendix
\section{A Subordinated Stochastic Process for the 5~minute
Returns} In section~2 the $T=5$~minute returns of a stock were
found to be well described by the pdf of\equ{specialpdf} with
$a=0$ and standard deviation $b=v_{T=5{\rm min}}$ (without loss in
generality, we consider only the case of vanishing mean return).
To characterize the directing process of a subordinated stochastic
process\cite{Bo60} that would give this pdf, consider the
following integral representation,
\ba{dir}
\frac{2 b^3}{\pi (x^2+b^2)^2}&=&\frac{2 b^3}{\pi}\int_0^\infty
\lambda d\lambda
e^{-\lambda (b^2+x^2)}\\
&=&\int_0^\infty\frac{ e^{1/(2 t)} dt}{\sqrt{2\pi t^5}}
\frac{\exp\left[-\frac{x^2}{2 b^2 t}\right]}{\sqrt{2\pi b^2 t}}\ .
\ea
One thus can interpret the pdf for the returns as the result of a
random walk with variance $t v_T^2$, where $t$ is itself a random
variable drawn from the positive distribution
\be{distv}
q(t)=\frac{ e^{1/(2 t)}}{\sqrt{2\pi t^5}}\ .
\ee
Clark interprets the stochastic process $\tau(t)$ whose increments
are drawn from the distribution $q(t)$ as "operational"
time\cite{Cl73}. Note that $q(t)$ has unit mean and infinite
variance and that the random variable $1/t$ is drawn from a
$\Gamma$ distribution. In fact, the probability for a large
variance, respectively operational time interval, itself falls off
as a power law with an exponent of $-(3/2)$.

Although the subordinate stochastic process described here
reproduces the observed distribution of returns for $T=5$min by
construction, it does not describe the observed time dependence of
this distribution: the temporal evolution is {\it not} shape
stable and a Gaussian fixed point is approached rather quickly.

\end{document}